\documentclass[useAMS,usenatbib]{mn2e}
\usepackage{amsmath}
\usepackage{amssymb}
\usepackage[dvips]{graphicx}

\newcommand{\cn}{\ensuremath{C_{\mathrm N}^2}}
\newcommand{\cnh}{\ensuremath{C_{\mathrm N}^2(h)}} 

\title[LOLAS: an optical turbulence profiler for the first kilometer]
{LOLAS: an optical turbulence profiler in the atmospheric boundary layer
 with extreme altitude-resolution.} 

\author[R. Avila et al.]{
R. Avila$^{1}$\thanks{E-mail: r.avila@astrosmo.unam.mx}; 
J. L. Avil{\'e}s$^{1,4}$;
R. W. Wilson$^{2}$;
M. Chun$^{3}$;
T. Butterley$^{2}$;
E. Carrasco$^{4}$;\\
$^{1}$Centro de Radioastronom{\'\i}a y Astrof{\'\i}sica, UNAM, Apartado
Postal 3-72, Morelia, Michoac\'an, C.P. 58089, M\'exico\\
$^{2}$University of Durham, Department of Physics, Rochester Building,
South Road, Durham DH1 3LE\\
$^{3}$Institute for Astronomy, University of Hawaii, 640 North A'ohoku
Place, 209 Hilo, Hawaii 96720-2700, USA\\
$^{4}$Instituto Nacional de Astrof{\'\i}sica, \'Optica y Electr\'onica, 
Luis Enrique Erro 1, Tonantzintla, Puebla, C.P. 72840,  M\'exico
}
\begin{document}

\date{}

\pagerange{\pageref{firstpage}--\pageref{lastpage}} \pubyear{2006}

\maketitle

\label{firstpage}

\begin{abstract}
We report the development and first results of an instrument called Low Layer
Scidar (LOLAS) which is aimed at the measurement of optical-turbulence profiles in the
atmospheric boundary layer with high altitude-resolution. The method is based on
the Generalized Scidar (GS) concept, but unlike the GS instruments which
need a 1-m or larger telescope, LOLAS is implemented on a dedicated
40-cm telescope, making it an independent 
instrument. The system is designed for widely separated double-star
targets, which enables the high altitude-resolution. Using a
$200^{\prime\prime}$-separation double-star, we have
obtained turbulence profiles with unprecedented  12-m resolution. 
 The system 
incorporates necessary novel algorithms for autoguiding, autofocus and image
stabilisation.  The results presented here were obtained at Mauna Kea
Observatory. They show LOLAS capabilities but cannot be considered as representative of the
site. A forthcoming paper will be devoted to the site characterisation.  The instrument was built as part of the Ground Layer
Turbulence Monitoring Campaign on Mauna Kea for Gemini Observatory.

\end{abstract}


\section{Introduction}
 \label{sec:intro}

In recent years, a number of efforts have been conducted in the domain of
astronomical adaptive optics to increase the field of view in which the
wave front is corrected from atmospheric-turbulence-induced
perturbations. Particularly, Ground Layer Adaptive Optics (GLAO)
\citep{Rig02,Tok04} is aimed
at compensating the wave front deformations arising from turbulence very
close to the ground and not compensating those originating at higher
altitudes. This idea is based on the facts that the compensation of lower
altitude turbulent layers provides wider corrected fields of view
\citep{c98} and that
turbulence close to the ground is generally the most intense (e.g.,
\citet{AMV+04}). 

To develop a GLAO system for a given site, it is required to have as much
knowledge as possible about the optical turbulence in the ground layer
(i.e. in the first kilometer of altitude above the site). For example,
\citet{LH06} recognize that the biggest unknown in their theoretical analysis of
the performance of an infrared GLAO system is the vertical structure of the
ground-layer turbulence (\cnh) and that proper measurements are urgent to validate
their results.

In view of such necessity, we have developed a method called Low Layer
SCIDAR (LOLAS), the concept of which was presented by \citet{AC04}. 
It is based on the Generalised SCIDAR method (GS) but uses a smaller
telescope and more widely separated double stars as targets. The method is explained in
\S~\ref{sec:concept}.
\citet{EM07} reported the use of the GS technique 
to obtain {\cn} profiles measurements with an altitude
resolution of a few tens of meters, which was the best GS resolution
achieved so far.  The fundamental difference of that work with the present one 
is that they use the 1.8-m Vatican Advanced Technology Telescope whereas we
use a 40-cm dedicated telescope, which makes of LOLAS an independent
monitor.  See \S~\ref{sec:results} for further
comments on the work by \citet{EM07}.

The experimental setup and data reduction for LOLAS is
presented in \S~\ref{sec:instrument}.  The first results are shown in
\S~\ref{sec:results} and finally, in \S~\ref{sec:conclusions} we give the
 conclusion of this work.   

\section[]{LOLAS Concept}
 \label{sec:concept}

LOLAS is based on the GS concept
\citep{AVM97,FTV98}, which consists of computing the 
normalised-mean spatial autocorrelation function of short exposure-time
images of the scintillation pattern
produced by a double star. The normalisation is performed with respect to
the autocorrelation of the mean image. The resulting normalised
autocorrelation map is dimensionally equivalent to a
scintillation index, which is dimensionless. Hereafter, we will refer to
this normalised-mean autocorrelation simply as the autocorrelation.

In the classical SCIDAR
\citep{RRV74,CAV87,Ver92}, the analysis is made in the pupil plane of the
telescope, which makes it insensitive to turbulence close to the
ground because the scintillation variance is proportional to $h^{5/6}$ 
\citep{Rod81a},
where $h$ is the altitude of the turbulent layer (acting as a phase screen).
In the GS the plane of the detector is made the conjugate of a virtual plane
(analysis plane) located at an altitude $h_\mathrm{gs}$ of the order of
  a few kilometers below the ground, hence $h_\mathrm{gs} < 0$. 
 In this case the distance relevant
for scintillation produced by a turbulent layer at an altitude $h$ is 
$|\,h-h_\mathrm{gs}\,|$, which makes the turbulence at ground level detectable.
Following \citet{AVM97} and references therein, the autocorrelation
function obtained can be written as 
\begin{eqnarray} 
C_\mathrm{gs}\left( \mathbf{r}\right)=
\int_{h_\mathrm{gs}}^{+\infty }{\rm d}h\;K\left(\mathbf{r} ,h-h_\mathrm{gs}\right) 
\;C_N^2\left(h\right) +N(\mathbf{r})\mbox{,}   
\label{eq:eqint}
\end{eqnarray}
where the kernel $K\left(\mathbf{r} ,h-h_\mathrm{gs}\right)$ is the theoretical autocorrelation function
produced by a single layer at an altitude $h$ with a unit $C_N^2$ and 
$N(\mathbf{r})$ is the measurement noise.  When $h<0$, {\cnh =0}. For a double star of angular
separation $\boldsymbol{\rho}$, the Kernel consists of three autocorrelation peaks: one
centered on the autocorrelation origin and the two others separated from
the first one by $\mathbf{d}_\mathrm{r}=\boldsymbol{\rho}|h -
h_\mathrm{gs}|$ and $\mathbf{d}_\mathrm{l}=-\boldsymbol{\rho}|h -
h_\mathrm{gs}|$. The altitude of the layer is determined from the  measurement of
  $\mathbf{d}_\mathrm{r}$ or  $\mathbf{d}_\mathrm{l}$ and the solution of
  the corresponding equation.

The altitude resolution is equal to $\Delta d/\rho$, where $\Delta d$ is
the minimal measurable difference of the position of two autocorrelation
peaks. The natural value of $\Delta d$ is the full width at half maximum
$L$ of the aucorrelation peaks: $L(h)=0.78\sqrt{\lambda (h-h_\mathrm{gs})}$ \citep{PDA01}, where
$\lambda$ is the wavelength. However, $\Delta d$ can be shorter than $L$ if
the inversion of Eq.~\ref{eq:eqint} is performed using a method that can
achieve  super-resolution like  Maximum Entropy or CLEAN. Both methods have 
been used in GS
measurements \citep{PDA01}. \citet{Fri95} analised the CLEAN algorithm and its implications
for super-resolution. Applying his results for GS leads to an altitude
resolution of
\begin{equation}
\Delta h=\frac{2}{3}\frac{L}{\rho}=0.52\frac{\sqrt{\lambda (h-h_\mathrm{gs})}}{\rho}.
\label{eq:Delta_h}
\end{equation}
 
\begin{figure}
\includegraphics[width=\columnwidth]{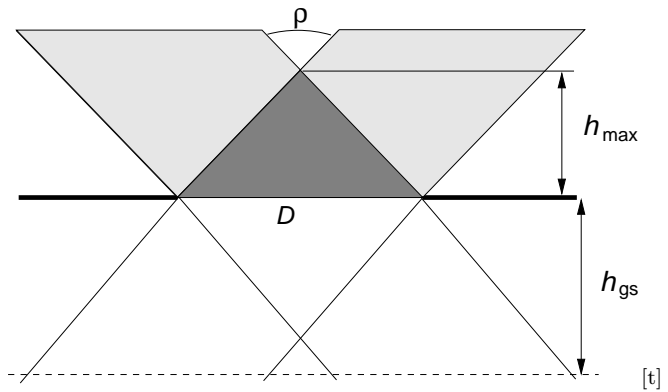}[t]
 \caption{Schematic for the determination of the maximum altitude 
$h_\mathrm{max}$ for which the {\cn} value can be
 retrieved. The altitude of the analysis plane $h_\mathrm{gs}$ is
 represented only to make clear that this value is not involved in the
 calculation of $h_\mathrm{max}$.}
 \label{fig:maxalt}
\end{figure}

The maximum altitude, $h_{\mathrm{max}}$ for which the {\cn} value 
 can be retrieved is set by the altitude at which the projections of the pupil
along the direction of each star cease to be overlapped, as no correlated
speckles would lie on the scintillation images coming from each
star.  Figure ~\ref{fig:maxalt} illustrates the basic geometrical
consideration involved in the determination of  $h_{\mathrm{max}}$ . 
Note that  $h_{\mathrm{max}}$ does not depend on $h_\mathrm{gs}$. The
 maximum altitude is thus given by
\begin{equation}
 h_{\mathrm{max}}=\frac{D}{\rho},  \label{maxalt}
\end{equation}
where $D$ is the pupil diameter.

In addition to the autocorrelation of the scintillation images, the GS 
calculates the mean cross-correlation of images taken at times separated by  a known
constant delay $\Delta t$. As explained by \citet{AVS01,PAD+04,ACI+06}, the
cross-correlation leads to the determination of the velocity of the
turbulent layers and the {\cn} of the turbulence generated inside the telescope dome.

LOLAS concept consists of the implementation of the GS technique on a
small dedicated telescope, using a very widely separated double
star. For example, for $h_\mathrm{gs}=-2$~km, $h=0$, 
$\lambda=0.5$~$\mu$m, $D=40$~cm and star separations of
180$^{\prime\prime}$ and 70$^{\prime\prime}$, $\Delta h$ equals 19 and
48~m, while  $h_\mathrm{max}$ equals 458 and 1179~m, respectively. 
GS uses a larger telescope (at least 1-m diameter) and closer double
stars, so that the entire altitude-range with non-negligible {\cn} values
is covered ($h_\mathrm{max}\gtrsim 30$~km). 
 
 The  altitude of the analysis plane, $h_\mathrm{gs}$
 was set to -2~km, as a result of a compromise between the increase of
 scintillation variance, which is proportional to $|h - h_\mathrm{gs}|^{5/6}$,
 and the reduction of pupil diffraction effects. Indeed, pupil diffraction
 caused by  the virtual distance between the pupil and the analysis planes
 provokes that  Eq.~\ref{eq:eqint} is only an approximation. The larger
 $h_\mathrm{gs}$  or the smaller the pupil diameter, the greater the error in applying 
Eq.~\ref{eq:eqint}. We have performed numerical simulations to estimate such
effect. 
A succinct description of the simulations is presented in Appendix
\ref{ap:sim}.

  The pixel size  projected on the pupil, $d_\mathrm{p}$, is set by the condition that the smallest scintillation speckles be sampled at the Nyquist spatial frequency
or better. The typical size of those speckles is equal to $L(0)$. Taking
the same values as above for $h_\mathrm{gs}$ and $\lambda$, yields $L(0)=2.45$~cm.
We chose $d_\mathrm{p}=1$~cm, which indeed satisfies the Nyquist criterion 
$d_\mathrm{p} \le L(0)/2$.
The altitude sampling of the turbulence profiles is  $\delta_h=d_p/\rho$. Note,
from the two last expressions and Eq.~\ref{eq:Delta_h}, that the altitude
resolution $\Delta h$ and the altitude sampling $\delta_h$ are related by 
$\delta_h \le (3/4) \Delta h$ for $h=0$.

\begin{figure}
\includegraphics[width=\columnwidth]{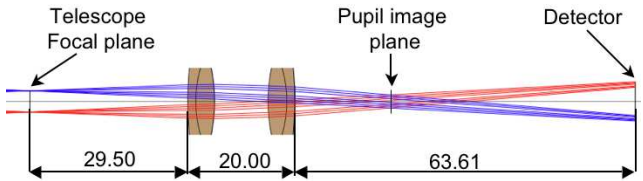}
 \caption{Optical layout. The dimensions unit is millimeters.}
        \label{fig:optics}
\end{figure}
\begin{figure}
\includegraphics[width=\columnwidth]{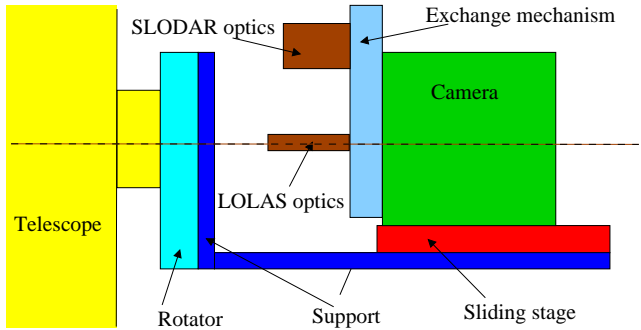}
 \caption{Schematic view of the instrument. The drawing is not on scale.}        \label{fig:schematics}
\end{figure}

\section{Instrument}
\label{sec:instrument}

\subsection{Hardware}
\label{sec:hardware}

A 40-cm Meade telescope installed on an equatorial mount sends light to
a pair of  50-mm focal-length achromatic lenses that form on the
detector an image of a virtual plane located approximately 2 km below the
pupil. Figure \ref{fig:optics} represents a scaled optical layout
showing the optical beams of two stars separated by 200$^{\prime\prime}$ as
they travel from the telescope focal plane to the detector. The beam
sizes correspond to the telescope-pupil diameter. 
The diameter of each beam at the 
detector plane is equal to 1.30~mm. In some GS systems a field lens
is used to shift the image of the analysis plane  either further away
from the telescope focus or closer to it, whether the lens is diverging or
converging, respectively. The optical setup for LOLAS was designed so to
avoid the use of a field lens.  
The detector is an Electron Multiplying Charged Couple
Device (EMCCD) manufactured by Andor Technology with 512$\times$512 square
pixels of 16 $\mu$m side each. The maker\footnote{www.andor-tech.com} asserts that
the effective readout noise is smaller than 1 electron r.m.s. and the
quantum efficiency is higher than 90\%. Frames are binned
2$\times$2. The size of each binned pixel on the conjugated plane is
9.8~mm.  One can chose any size for the active sub-array of the detector to
  be red.
 Figure~\ref{fig:schematics} shows a  schematic view of the
  instrument. The telescope holds a motrised rotator used to align the
  pixel lines along the separation of the double
star. The rotator holds the
  mechanical support that carries the rest of the instrument. A sliding
  stage, which enables precise positioning along the optical axis, is
  screwed on the support and carries the camera.  
The same equipment, except for the optics, is used for Slope
Detection and Ranging (SLODAR)
observations. The SLODAR \citep{WBG+04,BWS06} exploits Shack-Hartmann wavefront sensor 
measurements of the slope of the phase aberration produced by the atmospheric 
turbulence. This technique is implemented by replacing LOLAS focusing
optics by a collimator and a lenslet array. To easily switch between LOLAS
and SLODAR configurations, a manual mechanism is attached to the camera and
holds the optics of the two experiments. The mechanism is designed to
ensure the correct position of each optical assembly. 
 Figure~\ref{fig:schematics} shows the LOLAS optics barrel in
place.  Into this barrel the LOLAS focusing lenses are mounted. 
 The telescope, camera, stage and rotator are controlled by a Personal
Computer (PC) with two 3-GHz Xeon processors and 1-GB memory capacity,
running under Linux operating system. 

\begin{figure}
\includegraphics[width=\columnwidth]{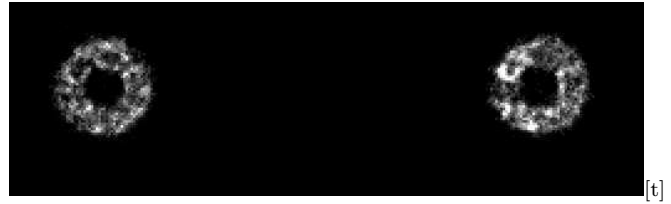}[t]
 \caption{Example of a scintillation image acquired with LOLAS. The
   exposure time was set to 3~ms.  The analysis plane is located at
   $h_\mathrm{gs}=-2$~km. The object is the double star  ``STF 4'' 
of coordinates
 $\alpha_{2000}=01\mathrm{h}\;56\rlap.^{\prime}2$ and
 $\delta_{2000}=+37^\circ\; 15^{\prime}$. The V-magnitudes of the stars are 5.68, 6.10 and their separation is $199\rlap.^{\prime\prime}7$.
        }
 \label{fig:pupils}
\end{figure}

\subsection{Data acquisition and processing}
\label{sec:data_acq_proc}

The out-of-focus pupil images produced by each star are centered on the
detector. To optimize the acquisition and processing speed, only a sub-frame of
256$\times$80 binned pixels is acquired. An example is shown in
Fig.~\ref{fig:pupils}. 
 The exposure time of each frame
ranges from 3 to 10 ms, depending on the wind conditions, and is selected
by the user. The typical number of frames used to obtain one set of auto-
and cross-correlations is 30000. The frame rate at 3~ms exposure time is
112 frames per second. 
 
In windy conditions, telescope shaking can cause
image motion which affects the calculation of the mean image. To avoid
this, every image is recentered prior to co-addition of the frames. In addition,
telescope tracking errors are corrected by updating the telescope position every
200 images, using the average image position of the most recently
acquired 200-frames packet. This autoguiding keeps the extra-pupil images on the active 
sub-array of the detector. It has been seen that the pupil images slowly
change in size. This is due to a slow shift of the telescope focus,
presumably caused by the redistribution of the primary mirror load while tracking the
stars. An autofocus system has been developed to overcome this problem. On
every average image calculated with 200 frames, the size of the out-of-focus pupils is
monitored and the camera and optics are moved by acting on the supporting
sliding stage, until the nominal size of 40 binned pixels is recovered.
Similarly, the alignment of the stars along the pixel lines is
 measured and eventually corrected automatically by acting on the
motorised rotator. Finally, the average flux over the pupil is also
monitored in each image and its variance relative to the mean flux is
calculated and stored in the header of the FITS file in which the
correlations are saved. This information is used to qualify the data. Indeed, mean
flux variations, due to cloud passages or fog condensation, are  adverse to
the retrieval of {\cn} profiles from the autocorrelations.

The auto- and cross- correlations are calculated as images are being
acquired. When a set of correlations are saved on disk, an IDL program is
automatically executed to invert Eq.~\ref{eq:eqint} and retrieve {\cnh}, using a modified
CLEAN algorithm  similar to that developed by \citet{PDA01} for GS measurements.

\section{First results}
 \label{sec:results}

The first results of LOLAS as described in
\S~\ref{sec:instrument} were obtained in September 2007 at Mauna Kea
Observatory, as part of a collaboration between the Universidad Nacional
Aut\'onoma de M\'exico (UNAM), the University of Durham (UD) and the
University of Hawaii (UH), under a contract with Gemini Observatory. The
instrument was installed on the Coud\'e roof of the UH 2.2-m telescope.  

\begin{figure}
\includegraphics[width=\columnwidth]{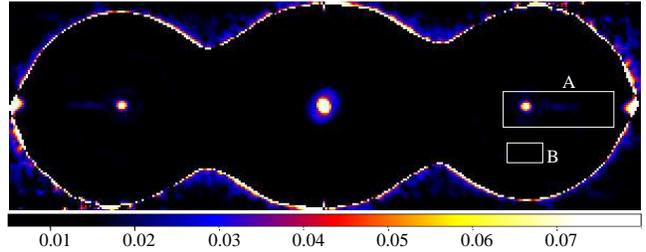}
 \caption{Example of an autocorrelation obtained with a moderately
   separated star ($\rho=108\rlap.^{\prime\prime}4$). The scale is
   indicated by the false-color code at the bottom. As explained in
   ~\S~\ref{sec:concept} the autocorrelations are dimensionless.  }       
 \label{fig:autocorr_close_bin}
\end{figure}

\begin{figure}
\includegraphics[width=\columnwidth]{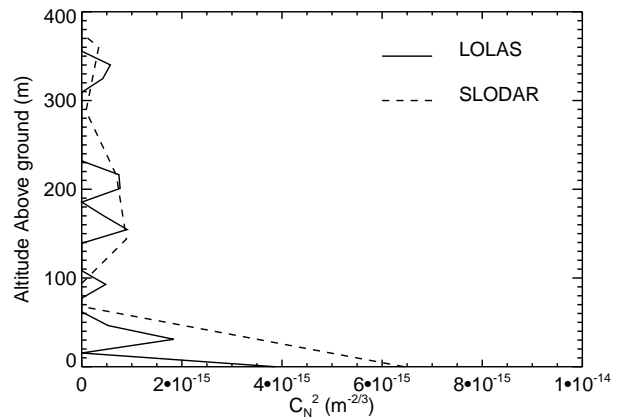} 
 \caption{Examples of {\cn} profiles. LOLAS profile was obtained from
   the autocorrelation shown in Fig.~\ref{fig:autocorr_close_bin}, in 2007
   July 3 at 9:51 UT. The SLODAR profile was obtained the same date at
   10:19 UT. {\cn} uncertainties for LOLAS and SLODAR are $\pm 1\times 10^{-16}$
   and  $\pm 3\times 10^{-16}$ m$^{-2/3}$.  }
 \label{fig:cn2-ls}
\end{figure}

\begin{figure}
\includegraphics[width=\columnwidth]{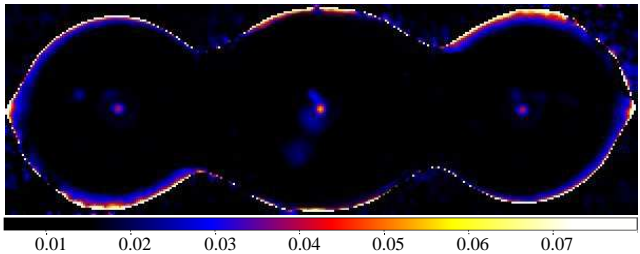}
 \caption{Cross correlation obtained with the same data as the
   autocorrelation shown in Fig.~\ref{fig:autocorr_close_bin}. }       
 \label{fig:crosscorr_close_bin}
\end{figure}

Figure~\ref{fig:autocorr_close_bin} shows an example of the autocorrelation
map of scintillation images obtained with  the double star ``STF 28'' of angular separation of 
$\rho=108\rlap.^{\prime\prime}4$, which is considered a moderate
separation for LOLAS. The number of frames used for this example was 30000 and the
exposure time of each frame was 3~ms.  The central peak is the result of
 the summation of the central peaks produced by all the turbulent layers in
 the atmosphere, even those that are higher than the maximum altitude $h_\mathrm{max}$
 reached by the instrument, which in this case is equal to 764~m (Eq.~\ref{maxalt}).
The correlated speckles produced by those layers are separated by a
longer distance than the pupil diameter, thereby not giving rise to lateral
peaks inside the autocorrelation-map bounds.   
Only the layers below $h_\mathrm{max}$ form visible lateral  peaks 
from which {\cnh} is retrieved. The right-hand-side wing of those peaks in
Fig.~\ref{fig:autocorr_close_bin} are framed by rectangle A. The
most intense peak inside that frame corresponds to the ground level and
is a consequence of the turbulence inside and outside the telescope tube.
Much fainter peaks further apart from the map center - thus from higher
layers - can be seen inside frame A.  It is worth noting that the
color scale for that figure was chosen such that the fainter peaks are
visible, not caring if the stronger peaks appear saturated. The
autocorrelation signal is contained inside a strip which is vertically
centered and aligned along the horizontal axis, because the detector pixels
were aligned along the direction of the double-star separation. Hence, the
values inside frame B correspond only to noise. The standard deviation
$\sigma_\mathrm{ac}$ of
those values is used as a measure of the noise level in the
data. In this case, $\sigma_\mathrm{ac}=4.5\times 10^{-4}$. 
Following \citet{Rod81a}, the relation between {\cn(0)} and a scintillation variance $\sigma_\mathrm{I}$ is:
\begin{equation}
\cn(0)=\frac{\sigma_\mathrm{I}\lambda^{7/6}|h_\mathrm{gs}|^{-5/6}}{19.12\;\delta_h}.
\end{equation}
For the data shown in Fig~\ref{fig:autocorr_close_bin}, $\delta_h=19$~m, hence $\sigma_\mathrm{ac}$ corresponds to a {\cn} uncertainty of $\Delta
\cn=1\times 10^{-16}$ m$^{-2/3}$.
Correlation
values larger than $3 \sigma_\mathrm{ac}$ are considered
as being actual signal. This sets the minimum detectable {\cn}
value to $C_\mathrm{N_{min}}^2=3\times10^{-16}$~$\mathrm{m}^{-2/3}$. Note
that $\Delta h$ is defined as the 
 resolution along the line-of-sight. For this case, $\Delta h= 31$~m, for
 h=0. The altitude resolution  at ground level in the
 vertical direction is  $\Delta h \cos(z)$, where $z$ is the zenith angle.
 In this case $z=35.6^\circ$ which results in a vertical resolution of 
   25.4~m.  The {\cn} profile corresponding to  that autocorrelation is shown in
 Fig.~\ref{fig:cn2-ls}. The contribution of turbulence inside the telescope
 tube has been removed from the {\cn} value at ground level. To do so,
 we used the temporal cross-correlation of images taken with a temporal
 interval of 13~ms (26 or 39~ms can also be chosen,
   depending on local wind speed), calculated with the same data as that used for the
 autocorrelation of Fig.~\ref{fig:autocorr_close_bin}. This
 cross-correlation is shown in
 Fig.~\ref{fig:crosscorr_close_bin}. The correlation triplet that is
 centered on the cross-correlation origin is interpreted as arising from
 static turbulence inside the telescope tube (see \citet{AVS01} for details
 of that method). The corresponding value of the {\cn} is 8.9$\times
 10^{-15}$~$\mathrm{m}^{-2/3}$, which is equivalent to
 $0\rlap.^{\prime\prime}37$ seeing.

In Fig.~\ref{fig:cn2-ls} is also shown a {\cn} profile measured with the
SLODAR, 28 minutes later than the measurement done with LOLAS, using the
same stellar source. The {\cn} uncertainty is obtained using the
bootstrap method.  
The 
altitude resolution achievable with LOLAS is approximately 5 times better than that 
for SLODAR. This is clearly illustrated by the profiles shown in
Fig.~\ref{fig:cn2-ls}: Around 200~m, LOLAS separates two layers, centered
at 150 and 210~m, while the SLODAR delivers a single {\cn} peak centered at
190~m, approximately. Similarly, in the 70~m, LOLAS identifies two
layers while the SLODAR only one. Note that the sum of the {\cn} values
delivered by LOLAS in the latter altitude range agrees with
the SLODAR value at ground level within 10\%. Both measurements are free of the
effect of the turbulence inside the telescope tube.  
 The lower altitude resolution of SLODAR is related to a wider spatial
sampling on the analysis plane. Here we used an array of $8\times 8$ sub-apertures,
each of diameter 5 cm, which is five times larger than $d_p$. This implies that  
the photon flux per spatial resolution element and hence the instantaneous 
signal-to-noise ratio of LOLAS measurements is  five times lower than for the SLODARs, 
so that a five times longer period of integration is required to measure the
turbulence profile to a given accuracy. 
Hence in practice a choice can be made between greater altitude  
resolution (LOLAS) or greater temporal resolution (SLODAR) of the profile.

\begin{figure}
\includegraphics[width=\columnwidth]{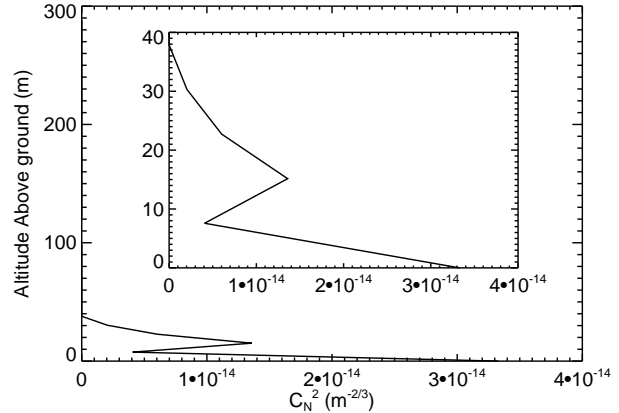}
 \caption{Example of a turbulence profile with the highest
   altitude-resolution so far obtained with LOLAS. The data was taken in
   2007 November 17 at 12:09 UT. The central frame shows an amplification
   of the profile in the low-altitude zone. The {\cn} uncertainty is
$\pm 1.6\times10^{-16}$ m$^{-2/3}$.}
 \label{fig:cn2_wide_bin}
\end{figure}

\begin{figure}
\includegraphics[width=\columnwidth]{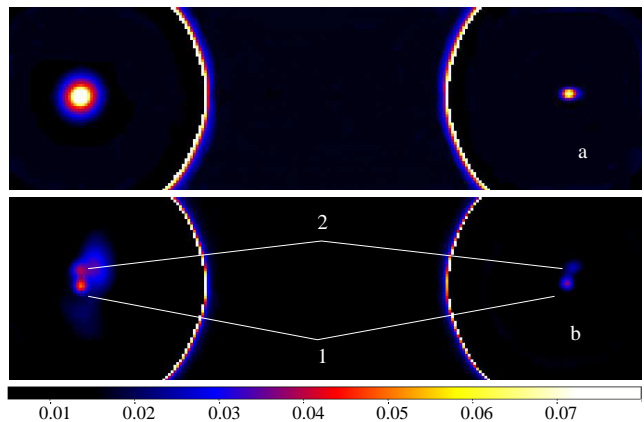}
 \caption{Similar to Figs.~\ref{fig:autocorr_close_bin} (for a) and
   \ref{fig:crosscorr_close_bin} (for b), but obtained with a widely
   separated double star ($\rho=199\rlap.^{\prime\prime}7$). }        
 \label{fig:ac_ic_wide_bin}
\end{figure}

To illustrate the highest altitude resolution that has so far been reached
with LOLAS, Fig. \ref{fig:cn2_wide_bin}
shows a {\cn} profile obtained using as target a
$199\rlap.^{\prime\prime}7$-separation double-star.  The altitude
  resolution in vertical direction is 11.7~m  and 
$\Delta \cn = 1.6\times10^{-16}\mathrm{m}^{-2/3}$. 
Note the ability for discerning a layer centered at 16 m from that at ground
layer. Similarly to the profile shown in Fig.~\ref{fig:cn2-ls}, the
turbulence inside the telescope tube has been removed. To make certain that
the {\cn} peak at 16 m is not an artifact produced by the inversion
algorithm, Fig.~\ref{fig:ac_ic_wide_bin} shows the corresponding
auto- and cross- correlations. Only an enlargement of the
central and right-hand-side peaks is shown. In the autocorrelation, the right-hand
side peak appears slightly elongated in the horizontal direction,
suggesting that this peak is in fact the result of two layers very close to
each other. This is confirmed on the cross-correlation map, where one can
clearly see the contribution of two distinct layers: peaks 1 correspond to
the ground layer and peaks 2 come from a slightly higher layer as the
separation of the latter peaks is larger than that of peaks 1.  The altitude
resolution might be improved if the information contained in the
cross-correlation map was systematically used \citep{EM07}.

\section{Conclusions}
 \label{sec:conclusions}

An instrument has been created for the monitoring of optical
turbulence profiles in the atmospheric boundary layer with an altitude
resolution that can reach 11.7~m. It is a stand-alone system that needs only an
external power supply, making it a suitable device for long-term monitoring
at developed and undeveloped astronomical sites. The system incorporates
algorithms that automatically preserve the correct
image position on the detector, focus and camera rotational position. 
Temporal variations of the pupil-averaged flux are also calculated 
automatically and stored in the output files as auxiliary data for data 
reduction. The computation of the
auto- and cross- correlation is performed in real time and the {\cn}
profile is automatically retrieved immediately after the correlations are 
stored on disk. Numerical simulations of the effect of pupil diffraction
show that it  reduces the values of the autocorrelation by
10\%. Nevertheless, this alteration is corrected before applying the
inversion algorithm.
The turbulence originated inside the telescope tube can
be removed from the {\cn} profile on a post-processing step. 
The first results have shown the very high capabilities of LOLAS in
terms of altitude resolution. The shortest   vertical resolution is
 only twice that of the instrumented balloons, which is of the order of
6~m \citep{AV05}. One disadvantage of LOLAS compared to the balloons, 
SLODAR, GS, Multi-Aperture Scintillation Sensor (MASS) \citep{KTV+03} or Single Star
Scidar (SSS) \citep{HVB+06} is that until now it does
not provide {\cn} values of the free atmosphere. However, scintillation
generated at higher layers is indeed detected with LOLAS. This information
could be used in two possible ways: performing a post-detection spatial
filtering using MASS-like masks or using SSS-like algorithms to retrieve \cn
values from the central peaks of cross-correlations obtained with a set of
different temporal intervals.     
SLODAR is the remote-sensing
technique that has a vertical resolution closer to that of LOLAS. The
advantage of SLODAR over LOLAS is that the temporal resolution of the {\cnh}
measurements is five times higher. A first
comparison of the profiles delivered by LOLAS and SLODAR have shown a
fairly good agreement between both techniques, although a more thorough
comparative study needs to be done to draw a definitive conclusion. The
turbulence profiles shown here are not to be taken as representative of the
site. A proper monitoring campaign at Mauna Kea is currently being carried
out and a statistical analysis of the results will be the subject of a
forthcoming paper.

\section*{Acknowledgments}

We are indebted to Marc Sarazin and ESO for their kind invitation and
funding to make use of the ESO SLODAR equipment for LOLAS very first
tests at Paranal, which proved to be a key experiment for the development
of the present instrument.  Data from the Washington Double Star
  Catalogue was of valuable help for this work. Funds for the instrument construction and
observations were provided by Gemini Observatory through contract number
0084699-GEM00445 entitled ``Contract for Ground Layer Turbulence Monitoring
Campaign on Mauna Kea''. Further funding was provided by grants IN111403
and IN112606-2 from DGAPA-UNAM.

%%%%%%%%%%%%%%%%%%%%%%%%%%%%%%%%%%%%%%%%%%%%%%%%%%%%%%%
%%%%% References %%%%%  
  
%\bibliographystyle{mn2e}%% MNRAS  
%\bibliography{/Users/remy/Lostex/ASTRONAT/mnras/mn-jour,/Users/remy/Lostex/BIBLIO/wholebib}

\begin{thebibliography}{}

\bibitem[\protect\citeauthoryear{Avila, Carrasco, Iba{\~n}ez, Vernin \&
  Cruz}{Avila et~al.}{2006}]{ACI+06}
Avila R.,  Carrasco E.,  Iba{\~n}ez F.,  Vernin J.,    Cruz D.,  2006, PASP,
  118, 503

\bibitem[\protect\citeauthoryear{Avila \& Chun}{Avila \& Chun}{2004}]{AC04}
Avila R.,  Chun M.,  2004, in Bonnaccini D.,  Ellerbroek B.,   Raggazzoni R.,
  eds, Advancements in Adaptive Optics Vol.~5490, A method for high altitude
  resolution $c_\mathrm{N}^2$ profiling in the first few hundred meters.
pp 742--748

\bibitem[\protect\citeauthoryear{Avila, Masciadri, Vernin \& S{\'anchez}}{Avila
  et~al.}{2004}]{AMV+04}
Avila R.,  Masciadri E.,  Vernin J.,    S{\'anchez} L.,  2004, PASP, 116, 682

\bibitem[\protect\citeauthoryear{Avila, Vernin \& Masciadri}{Avila
  et~al.}{1997}]{AVM97}
Avila R.,  Vernin J.,    Masciadri E.,  1997, Appl. Opt., 36, 7898

\bibitem[\protect\citeauthoryear{Avila, Vernin \& S\'anchez}{Avila
  et~al.}{2001}]{AVS01}
Avila R.,  Vernin J.,    S\'anchez L.~J.,  2001, A\&A, 369, 364

\bibitem[\protect\citeauthoryear{Azouit \& Vernin}{Azouit \&
  Vernin}{2005}]{AV05}
Azouit M.,  Vernin J.,  2005, PASP, 117, 536

\bibitem[\protect\citeauthoryear{Butterley, Wilson \& Sarazin}{Butterley
  et~al.}{2006}]{BWS06}
Butterley T.,  Wilson R.~W.,    Sarazin M.,  2006, 369, 835

\bibitem[\protect\citeauthoryear{Caccia, Azouit \& Vernin}{Caccia
  et~al.}{1987}]{CAV87}
Caccia J.~L.,  Azouit M.,    Vernin J.,  1987, Appl. Opt., 26, 1288

\bibitem[\protect\citeauthoryear{Chun}{Chun}{1998}]{c98}
Chun M.,  1998, PASP, 110, 317

\bibitem[\protect\citeauthoryear{Cruz, Gonz{\'a}lez, Angeles, Avila,
  S{\'a}nchez, Iriarte, Cuevas, Mart{\'\i}nez, Farah, S{\'a}nchez \&
  Mart{\'\i}nez}{Cruz et~al.}{2003}]{CGA+03}
Cruz D.~X.,  Gonz{\'a}lez S.~I.,  Angeles F.,  Avila R.,  S{\'a}nchez L.~J.,
  Iriarte A.,  Cuevas S.,  Mart{\'\i}nez L.~A.,  Farah A.,  S{\'a}nchez B.,
  Mart{\'\i}nez M.,  2003, in Cruz-Gonz{\'a}lez I.,  Avila R.,   Tapia M.,
  eds, San Pedro M{\'a}rtir: Astronomical Site Evaluation Vol.~19, Development
  of a generalized scidar at unam.
Revista Mexicana de Astronom{\'\i}a y Astrof{\'\i}sica (serie de conferencias),
  pp 44--51

\bibitem[\protect\citeauthoryear{Egner \& Masciadri}{Egner \&
  Masciadri}{2007}]{EM07}
Egner S.~E.,  Masciadri E.,  2007, PASP, 119, 1441

\bibitem[\protect\citeauthoryear{Fried}{Fried}{1995}]{Fri95}
Fried D.~L.,  1995, J. Opt. Soc. Am. A, 12, 853

\bibitem[\protect\citeauthoryear{Fuchs, Tallon \& Vernin}{Fuchs
  et~al.}{1998}]{FTV98}
Fuchs A.,  Tallon M.,    Vernin J.,  1998, PASP, 110, 86

\bibitem[\protect\citeauthoryear{Habbib, Vernin, Benkhaldoun \& Lanteri}{Habbib
  et~al.}{2006}]{HVB+06}
Habbib A.,  Vernin J.,  Benkhaldoun Z.,    Lanteri H.,  2006, MNRAS, 368, 1456

\bibitem[\protect\citeauthoryear{Kornilov, A.Tokovinin, Vozyakova, A., Shatsky,
  S. \& Sarazin}{Kornilov et~al.}{2003}]{KTV+03}
Kornilov V.,  A.Tokovinin Vozyakova O.,  A. Shatsky N.,  S. S.~P.,    Sarazin
  M.,  2003, in Wizinowich P.,  Bonnaccini D.,  eds, Adaptive Optical System
  Technologies II Vol.~4839, Mass: a monitor of the vertical turbulence
  distribution.
pp 837--845

\bibitem[\protect\citeauthoryear{{Le~Louarn} \& Hubin}{{Le~Louarn} \&
  Hubin}{2006}]{LH06}
{Le~Louarn} M.,  Hubin N.,  2006, MNRAS, 365, 1324

\bibitem[\protect\citeauthoryear{Prieur, Avila, Daigne \& Vernin}{Prieur
  et~al.}{2004}]{PAD+04}
Prieur J.~L.,  Avila R.,  Daigne G.,    Vernin J.,  2004, PASP, 116, 778

\bibitem[\protect\citeauthoryear{Prieur, Daigne \& Avila}{Prieur
  et~al.}{2001}]{PDA01}
Prieur J.-L.,  Daigne G.,    Avila R.,  2001, A\&A, 371, 366

\bibitem[\protect\citeauthoryear{Rigaut}{Rigaut}{2002}]{Rig02}
Rigaut F.,  2002, in Vernet E.,  Ragazzoni R.,  Esposito S.,   Hubin N.,  eds,
  Beyond conventional adaptive optics Vol.~58, 3d optical turbulence
  characterization for the new class of adaptive optics techniques.
ESO Conference and Workshop Proceedings, pp 11--16

\bibitem[\protect\citeauthoryear{Rocca, Roddier \& Vernin}{Rocca
  et~al.}{1974}]{RRV74}
Rocca A.,  Roddier F.,    Vernin J.,  1974, J. Opt. Soc. Am. A, 64, 1000

\bibitem[\protect\citeauthoryear{Roddier}{Roddier}{1981}]{Rod81a}
Roddier F.,  1981, Progress in Optics, XIX, 281

\bibitem[\protect\citeauthoryear{Tokovinin}{Tokovinin}{2004}]{Tok04}
Tokovinin A.,  2004, PASP, 116, 941

\bibitem[\protect\citeauthoryear{Vernin}{Vernin}{1992}]{Ver92}
Vernin J.,  1992, in Tatarskii V.~I.,  Ishimaru A.,   Zavorotny V.~U.,  eds,
  Wave propagation in Random Media (Scintillation) - Invited papers Atmospheric
  turbulence profiles.
SPIE Press, Bellingham, Wash., pp 248--260

\bibitem[\protect\citeauthoryear{Wilson, Bate, Guerra, Sarazin \&
  Saunter}{Wilson et~al.}{2004}]{WBG+04}
Wilson R.~W.,  Bate J.,  Guerra J.~C.,  Sarazin M.,    Saunter C.,  2004, in
  Bonnaccini D.,  Ellerbroek B.,   Raggazzoni R.,  eds, Advancements in
  Adaptive Optics Vol.~5490, Development of a portable slodar turbulence
  profiler.
pp 758--765

\end{thebibliography}

\appendix  

\section{Pupil diffraction effect}
\label{ap:sim}

The effect of pupil diffraction has been investigated by numerical
simulations of the normalised autocorrelation of scintillation images of a
double star, on a 40-cm telescope.  There is no analytical investigation of
this effect in the literature.
We used the SCIDAR Simulator \citep{CGA+03} which is based on the
optical turbulence simulator named
Turbulenz\footnote{http://www.mpia.de/AO/ATMOSPHERE/TurbuLenZ/tlz.html}.

Plane waves coming from two stars separated by 1 arc-minute pass
through three turbulent layers located at altitudes $h$ of 100, 300 and
700~m above the ground. The three layers have equal {\cn} corresponding to
a Fried parameter of 20 cm.
 The phase of the waves are distorted by the turbulent layers
following a Kolmogorov spectrum. Between the layers and down to 
the simulated detection plane, wave propagation is taken into account by
calculating Fresnel diffraction of the complex amplitudes. At ground level,
a mask is applied to the complex amplitudes, which simulates an unaberrated
pupil of a 40-cm Meade telescope, including the central obscuration. 
The intensity of the resulting waves are calculated on the detection plane,
located 2 km below the ground (i.e. $h_\mathrm{gs}=-2$ km). 
The autocorrelation of the intensity map is
calculated and co-added for 1000 statistically independent samples. This
co-added autocorrelation is normalized by the autocorrelation of the mean 
intensity. The amplitude and shape of the correlation peaks obtained for
the corresponding  layers are compared to the theoretical ones that would be
obtained with an infinite pupil-size.  Those theoretical values are
calculated by considering that the correlation amplitude is proportional to
$\cn h^{5/6}$ and the width of the correlation peak is
proportional to $\sqrt{\lambda L}$. The amplitude of the simulated
correlation peaks are 3.4\%, 8.4\% and 9.3\% smaller than the amplitudes of
the theoretical peaks, for the layers at 100, 300 and 700 m, respectively.
The measured autocorrelation peaks are corrected by a factor obtained by
the interpolation of those factors for the altitude of the corresponding
layer, before applying the inversion procedure which is designed for an
infinite pupil-size.  
\label{lastpage}

\end{document}